\documentclass[aps,prl,twocolumn,superscriptaddress,showpacs,letterpaper]{revtex4}
\usepackage{multirow}
\usepackage{color}
\usepackage{eso-pic} 
\usepackage{fancyhdr}
\usepackage{longtable}
\usepackage{rotating}
\usepackage{amssymb} 
\usepackage{amsmath}
\RequirePackage{setspace}
\usepackage{axodraw}
\usepackage{verbatim}

\usepackage{graphicx}% Include figure files
\usepackage{dcolumn}% Align table columns on decimal point
\usepackage{bm}% bold math

\usepackage{amsmath}
\usepackage{mathrsfs}
\usepackage{hyperref}

\newcommand{\Met}{\mbox{${ \not\!\!E}_{T}$}}

\newcommand{\MetVec}{\mbox{${\vec{ \not\!\!E}}_{T}$}}

\begin{document}

% remove the space for publication
\title{Search for $WW$ and $WZ$ production in lepton plus jets final state at CDF}

\affiliation{Institute of Physics, Academia Sinica, Taipei, Taiwan 11529, Republic of China} 
\affiliation{Argonne National Laboratory, Argonne, Illinois 60439} 
\affiliation{University of Athens, 157 71 Athens, Greece} 
\affiliation{Institut de Fisica d'Altes Energies, Universitat Autonoma de Barcelona, E-08193, Bellaterra (Barcelona), Spain} 
\affiliation{Baylor University, Waco, Texas  76798} 
\affiliation{Istituto Nazionale di Fisica Nucleare Bologna, $^x$University of Bologna, I-40127 Bologna, Italy} 
\affiliation{Brandeis University, Waltham, Massachusetts 02254} 
\affiliation{University of California, Davis, Davis, California  95616} 
\affiliation{University of California, Los Angeles, Los Angeles, California  90024} 
\affiliation{University of California, San Diego, La Jolla, California  92093} 
\affiliation{University of California, Santa Barbara, Santa Barbara, California 93106} 
\affiliation{Instituto de Fisica de Cantabria, CSIC-University of Cantabria, 39005 Santander, Spain} 
\affiliation{Carnegie Mellon University, Pittsburgh, PA  15213} 
\affiliation{Enrico Fermi Institute, University of Chicago, Chicago, Illinois 60637}
\affiliation{Comenius University, 842 48 Bratislava, Slovakia; Institute of Experimental Physics, 040 01 Kosice, Slovakia} 
\affiliation{Joint Institute for Nuclear Research, RU-141980 Dubna, Russia} 
\affiliation{Duke University, Durham, North Carolina  27708} 
\affiliation{Fermi National Accelerator Laboratory, Batavia, Illinois 60510} 
\affiliation{University of Florida, Gainesville, Florida  32611} 
\affiliation{Laboratori Nazionali di Frascati, Istituto Nazionale di Fisica Nucleare, I-00044 Frascati, Italy} 
\affiliation{University of Geneva, CH-1211 Geneva 4, Switzerland} 
\affiliation{Glasgow University, Glasgow G12 8QQ, United Kingdom} 
\affiliation{Harvard University, Cambridge, Massachusetts 02138} 
\affiliation{Division of High Energy Physics, Department of Physics, University of Helsinki and Helsinki Institute of Physics, FIN-00014, Helsinki, Finland} 
\affiliation{University of Illinois, Urbana, Illinois 61801} 
\affiliation{The Johns Hopkins University, Baltimore, Maryland 21218} 
\affiliation{Institut f\"{u}r Experimentelle Kernphysik, Universit\"{a}t Karlsruhe, 76128 Karlsruhe, Germany} 
\affiliation{Center for High Energy Physics: Kyungpook National University, Daegu 702-701, Korea; Seoul National University, Seoul 151-742, Korea; Sungkyunkwan University, Suwon 440-746, Korea; Korea Institute of Science and Technology Information, Daejeon, 305-806, Korea; Chonnam National University, Gwangju, 500-757, Korea} 
\affiliation{Ernest Orlando Lawrence Berkeley National Laboratory, Berkeley, California 94720} 
\affiliation{University of Liverpool, Liverpool L69 7ZE, United Kingdom} 
\affiliation{University College London, London WC1E 6BT, United Kingdom} 
\affiliation{Centro de Investigaciones Energeticas Medioambientales y Tecnologicas, E-28040 Madrid, Spain} 
\affiliation{Massachusetts Institute of Technology, Cambridge, Massachusetts  02139} 
\affiliation{Institute of Particle Physics: McGill University, Montr\'{e}al, Qu\'{e}bec, Canada H3A~2T8; Simon Fraser University, Burnaby, British Columbia, Canada V5A~1S6; University of Toronto, Toronto, Ontario, Canada M5S~1A7; and TRIUMF, Vancouver, British Columbia, Canada V6T~2A3} 
\affiliation{University of Michigan, Ann Arbor, Michigan 48109} 
\affiliation{Michigan State University, East Lansing, Michigan  48824}
\affiliation{Institution for Theoretical and Experimental Physics, ITEP, Moscow 117259, Russia} 
\affiliation{University of New Mexico, Albuquerque, New Mexico 87131} 
\affiliation{Northwestern University, Evanston, Illinois  60208} 
\affiliation{The Ohio State University, Columbus, Ohio  43210} 
\affiliation{Okayama University, Okayama 700-8530, Japan} 
\affiliation{Osaka City University, Osaka 588, Japan} 
\affiliation{University of Oxford, Oxford OX1 3RH, United Kingdom} 
\affiliation{Istituto Nazionale di Fisica Nucleare, Sezione di Padova-Trento, $^y$University of Padova, I-35131 Padova, Italy} 
\affiliation{LPNHE, Universite Pierre et Marie Curie/IN2P3-CNRS, UMR7585, Paris, F-75252 France} 
\affiliation{University of Pennsylvania, Philadelphia, Pennsylvania 19104}
\affiliation{Istituto Nazionale di Fisica Nucleare Pisa, $^z$University of Pisa, $^{aa}$University of Siena and $^{bb}$Scuola Normale Superiore, I-56127 Pisa, Italy} 
\affiliation{University of Pittsburgh, Pittsburgh, Pennsylvania 15260} 
\affiliation{Purdue University, West Lafayette, Indiana 47907} 
\affiliation{University of Rochester, Rochester, New York 14627} 
\affiliation{The Rockefeller University, New York, New York 10021} 
\affiliation{Istituto Nazionale di Fisica Nucleare, Sezione di Roma 1, $^{cc}$Sapienza Universit\`{a} di Roma, I-00185 Roma, Italy} 

\affiliation{Rutgers University, Piscataway, New Jersey 08855} 
\affiliation{Texas A\&M University, College Station, Texas 77843} 
\affiliation{Istituto Nazionale di Fisica Nucleare Trieste/Udine, I-34100 Trieste, $^{dd}$University of Trieste/Udine, I-33100 Udine, Italy} 
\affiliation{University of Tsukuba, Tsukuba, Ibaraki 305, Japan} 
\affiliation{Tufts University, Medford, Massachusetts 02155} 
\affiliation{Waseda University, Tokyo 169, Japan} 
\affiliation{Wayne State University, Detroit, Michigan  48201} 
\affiliation{University of Wisconsin, Madison, Wisconsin 53706} 
\affiliation{Yale University, New Haven, Connecticut 06520} 
\author{T.~Aaltonen}
\affiliation{Division of High Energy Physics, Department of Physics, University of Helsinki and Helsinki Institute of Physics, FIN-00014, Helsinki, Finland}
\author{J.~Adelman}
\affiliation{Enrico Fermi Institute, University of Chicago, Chicago, Illinois 60637}
\author{T.~Akimoto}
\affiliation{University of Tsukuba, Tsukuba, Ibaraki 305, Japan}
\author{B.~\'{A}lvarez~Gonz\'{a}lez$^s$}
\affiliation{Instituto de Fisica de Cantabria, CSIC-University of Cantabria, 39005 Santander, Spain}
\author{S.~Amerio$^y$}
\affiliation{Istituto Nazionale di Fisica Nucleare, Sezione di Padova-Trento, $^y$University of Padova, I-35131 Padova, Italy} 

\author{D.~Amidei}
\affiliation{University of Michigan, Ann Arbor, Michigan 48109}
\author{A.~Anastassov}
\affiliation{Northwestern University, Evanston, Illinois  60208}
\author{A.~Annovi}
\affiliation{Laboratori Nazionali di Frascati, Istituto Nazionale di Fisica Nucleare, I-00044 Frascati, Italy}
\author{J.~Antos}
\affiliation{Comenius University, 842 48 Bratislava, Slovakia; Institute of Experimental Physics, 040 01 Kosice, Slovakia}
\author{G.~Apollinari}
\affiliation{Fermi National Accelerator Laboratory, Batavia, Illinois 60510}
% ...

\collaboration{CDF Collaboration\footnote{With visitors from $^a$University of Massachusetts Amherst, Amherst, Massachusetts 01003,
$^b$Universiteit Antwerpen, B-2610 Antwerp, Belgium, 
$^c$University of Bristol, Bristol BS8 1TL, United Kingdom,
$^d$Chinese Academy of Sciences, Beijing 100864, China, 
$^e$Istituto Nazionale di Fisica Nucleare, Sezione di Cagliari, 09042 Monserrato (Cagliari), Italy,
$^f$University of California Irvine, Irvine, CA  92697, 
$^g$University of California Santa Cruz, Santa Cruz, CA  95064, 
$^h$Cornell University, Ithaca, NY  14853, 
$^i$University of Cyprus, Nicosia CY-1678, Cyprus, 
$^j$University College Dublin, Dublin 4, Ireland,
$^k$University of Edinburgh, Edinburgh EH9 3JZ, United Kingdom, 
$^l$University of Fukui, Fukui City, Fukui Prefecture, Japan 910-0017
$^m$Kinki University, Higashi-Osaka City, Japan 577-8502
$^n$Universidad Iberoamericana, Mexico D.F., Mexico,
$^o$Queen Mary, University of London, London, E1 4NS, England,
$^p$University of Manchester, Manchester M13 9PL, England, 
$^q$Nagasaki Institute of Applied Science, Nagasaki, Japan, 
$^r$University of Notre Dame, Notre Dame, IN 46556,
$^s$University de Oviedo, E-33007 Oviedo, Spain, 
$^t$Texas Tech University, Lubbock, TX  79609, 
$^u$IFIC(CSIC-Universitat de Valencia), 46071 Valencia, Spain,
$^v$University of Virginia, Charlottesville, VA  22904,
$^w$Bergische Universit\"at Wuppertal, 42097 Wuppertal, Germany,
$^{ee}$On leave from J.~Stefan Institute, Ljubljana, Slovenia, 
}}
\noaffiliation

\begin{abstract}

We present a search for $WW$ and $WZ$ production in final states that contain a charged lepton (electron or muon) and at least two jets, produced in $\sqrt{s} = 1.96~\rm{TeV}$ $p\bar{p}$ collisions at the Fermilab Tevatron, using data corresponding to 1.2 fb$^{-1}$ of integrated luminosity collected with the CDF II
detector. Diboson production in this decay channel has yet to be observed at hadron colliders due to the large
single $W$ plus jets background. An artificial neural network has been developed to increase signal sensitivity, as compared with an event selection based on conventional cuts. We set a 95\% confidence level upper limit of $\sigma_{WW}\times BR(W\rightarrow \ell\nu_{\ell},W\rightarrow {\rm jets})+\sigma_{WZ}\times BR(W\rightarrow \ell\nu_{\ell},Z\rightarrow {\rm jets} ) <2.88~\rm{pb}$, which is consistent with the standard model next to leading order cross section calculation for this decay channel of $2.09\pm0.12~\rm{pb}$.
\\

\end{abstract}

% activate the following line for publication
% \pacs{Valid PACS appear here}

\maketitle

%%%%%%%%%%%%%%%%%%%%%%%%%%%%%%%%%%%%%%
%%\section{\label{sec:Intro}Introduction}
%%%%%%%%%%%%%%%%%%%%%%%%%%%%%%%%%%%%%%

\begin{figure}[h]
  \begin{center}
    \includegraphics[trim = 55mm 180mm 40mm 40mm, clip, width=0.48\textwidth]{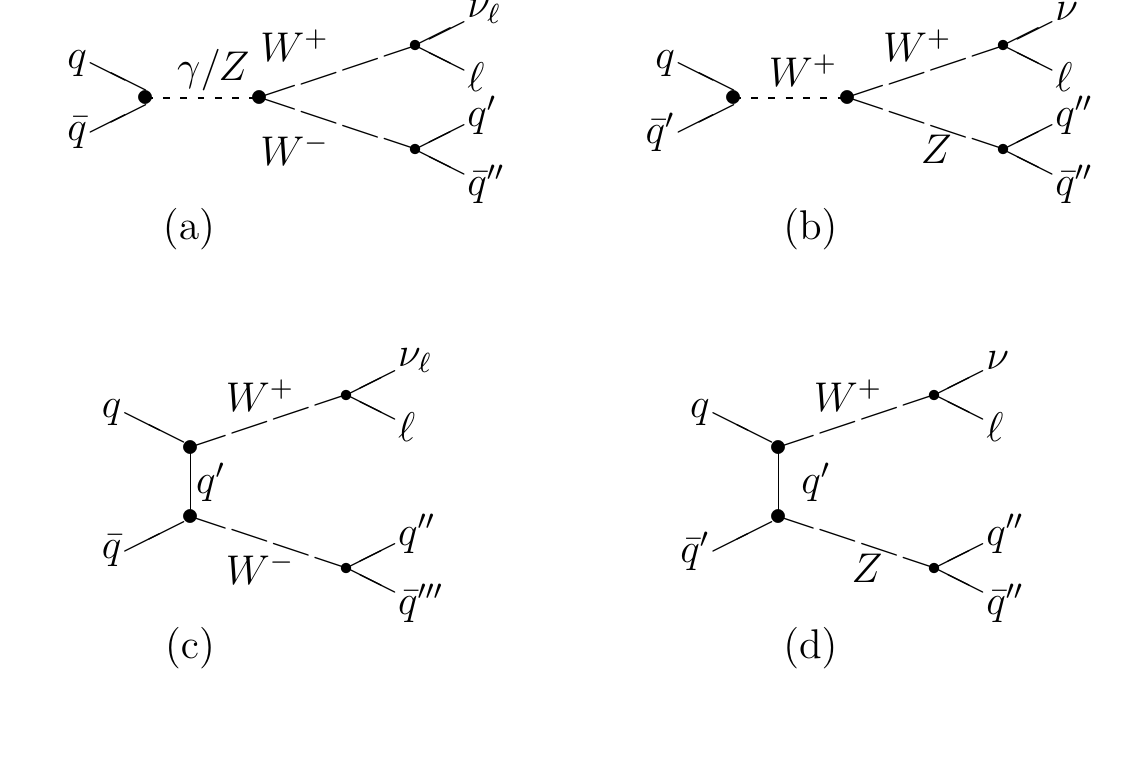}
\caption{Leading-order diagrams for $WW$ and $WZ$ production in the semi-leptonic decay channel; there are similar diagrams for the charge conjugates of the processes shown.}
\label{Fig:feyn_decay} 
  \end{center}
\end{figure}

In the standard model of particle physics (SM), the weak bosons ($W,~Z$) and the photon are the gauge bosons of the local $SU(2)~\times~U(1)$ symmetry. The spontaneous breaking of this symmetry gives masses to the $W$ and $Z$ bosons, while the gauge symmetry itself defines the interactions among these heavy bosons and the photon. Since the electroweak sector of the SM relies on this mechanism, it is of prime importance to test the boson couplings experimentally. We present in this paper a search for $WW$ and $WZ$ production in the charged lepton (electron or muon), neutrino plus jets decay channel \cite{my_thesis}. Figure \ref{Fig:feyn_decay} shows the leading order diagrams for the $p{\bar p}\rightarrow W(\rightarrow \ell \nu) V(\rightarrow {\rm jets})$ process, where $V\equiv W, Z$.  
      
The production of $WW$ and $WZ$ could be more sensitive to the triple gauge couplings (TGC) $WW(Z/\gamma)$, present in the s-channel (Fig. \ref{Fig:feyn_decay}a, b), and would be enhanced by the presence of nonstandard couplings (anomalous TGC) \cite{WWZ}. The hadronically decaying $W$ ($W\rightarrow \rm{jets}$) cannot be differentiated from hadronically decaying $Z$ ($Z\rightarrow \rm{jets}$) due to the limited jet energy resolution of the detector \cite{corrections}. We therefore search for the combined $WW$ and $WZ$ production.

The next to leading order (NLO) SM cross sections times branching ratio for these modes are $\sigma_{WW}\times BR(W\rightarrow \ell\nu_{\ell},W\rightarrow \rm{jets}) = 1.81\pm0.12~\rm{pb}$ and $\sigma_{WZ}\times BR(W\rightarrow \ell\nu_{\ell},Z\rightarrow \rm{jets} )  = 0.28\pm0.02~\rm{pb}$ \cite{pdg,Ellis_theory}.

The D0 collaboration recently reported the first evidence for the $WW$ and $WZ$ production in lepton plus jets decay mode \cite{D0paper}. This decay mode has not been observed yet at hadron colliders due to the large $W$ plus jets background. The cross section for $W$ plus two or more jets production at $\sqrt{s}=1.96$~TeV is at least two orders of magnitude larger than the total cross section times branching ratio of the signal \cite{messina}, which translates into an expected signal to background ratio of less than 1$\%$. Given that this diboson production is topologically similar to the associated production of Higgs and $W$ bosons, techniques that are developed for the $WW$ and $WZ$ searches are of key importance for Higgs searches. The lepton plus jets final state is common in other interesting processes, such as top production; thus, diboson production decaying in this channel is a significant background to these processes, and vice-versa.

%%\section{\label{sec:data_events}Event Selection}

The search for $WW$ and $WZ$ production is performed using data corresponding to 1.2 fb$^{-1}$ of integrated luminosity collected with the CDF II detector from $p\bar{p}$ collisions at $\sqrt{s} = 1.96~\rm{TeV}$ at the Fermilab Tevatron. The CDF II detector  is a general-purpose, multilayered detector designed to study many aspects of particle physics. It combines precise tracking systems with calorimeters and muon detectors \cite{CDF_Detector}. A tracking system is positioned closest to the beamline to provide accurate momentum determination of charged particles. The tracking system is immersed in a 1.4~T uniform magnetic field, produced by a superconducting solenoid and aligned along the proton direction. Calorimeters located outside the tracking volume provide energy measurement of electrons, photons, and jets. The geometrical coverage of the calorimeters is maximized to measure the energy flow of all particles produced in a collision and indirectly detect the neutrinos by the presence of missing transverse energy $\Met$ \cite{definitions}. Muon chambers are located on the outer part of the CDF II detector. 

The selection of the signal events proceeds as follows. The trigger system selects events with leptons of central pseudorapidity $|\eta|<1$; electron candidates with transverse energy $E_T >18~\rm{GeV}$ or muon candidates with transverse momentum $p_T>18~\rm{GeV/}c$. Events that are reconstructed offline are required to contain one electron candidate with $E_T>25~\rm{GeV}$ or one muon candidate with $p_T>25~\rm{GeV/}c$. The sample is enriched in events containing a neutrino by requiring that the $\Met$, corrected for the calorimeter energy leakage and the presence of muons, satisfies $\Met>25~\rm{GeV}$.

The jets are reconstructed in the calorimeter using the {\sc jetclu} cone algorithm \cite{jets} with cone radius $R=\sqrt{\Delta\phi^2+\Delta\eta^2}=0.4$. Starting from seed locations corresponding to calorimeter towers with $E_T>1~\rm{GeV}$, all nearby towers with $E_T>0.1~\rm{GeV}$ are used to search for stable cones. To resolve ambiguities due to overlaps, cones sharing an energy fraction greater than 0.75 are merged into a single jet. The measured energy deposition in the detector is corrected for effects that distort the true jet energy \cite{corrections}. Such effects include the non-linear response of the calorimeter to the particle energy, uninstrumented regions of the detector, spectator interactions, and energy radiated outside the cone. We select events that contain two or more jets with $E_T ^{jet}>15~\rm{GeV}$. To enhance the signal selection, events are rejected if the difference in pseudorapidity between the two leading jets, $\Delta \rm{\eta}(Jet1,Jet2)$, is greater than 2.5.

The $WW$ and $WZ$ production in the lepton plus jets event signature is simulated using {\sc pythia} v6.3 \cite{pythia} Monte Carlo generator, followed by {\sc geant}-based \cite{geant} CDF detector simulation. We search for diboson production in the region of [45,160]~GeV/$c^2$ in the dijet invariant mass, that is constructed taking the two leading jets into account. Using the signal Monte Carlo description, we define a signal region of [60,100]~GeV/$c^2$. It contains approximately 80\% of the reconstructed hadronically decaying $W$ bosons. Outside of the signal region we define a lower sideband region of [45,60]~GeV/$c^2$ and a higher sideband region of [100,160]~GeV/$c^2$. We enhance the $W$ event selection by rejecting events if the transverse mass $M_T$ of the lepton and $\Met$ system is not within the interval $30~{\rm GeV/c^2}<M_T<120~{\rm GeV/}c^2$. 

The most significant background to the $WW$ and $WZ$ search in the lepton plus jets decay channel consists of $W$ plus jets events where the leptonically decaying $W$ boson is produced in association with jets that mimic a hadronically decaying $W$ or $Z$. The $W$ plus jets background is simulated using the {\sc alpgen} v1.3 \cite{alpgen} Monte Carlo generator, followed by {\sc pythia} Monte Carlo generator for the parton shower and fragmentation, and full {\sc geant} detector simulation. Other, less significant backgrounds originate from a tau lepton that is detected as an electron or a muon; events with large transverse energy due to  the Drell-Yan process, where one of the two leptons is not reconstructed; and QCD events with a jet misidentified as a lepton. The QCD background is derived from the data, while the other background processes are simulated using {\sc pythia} Monte Carlo events.

%%%%%%%%%%%%%%%
Based on the Monte Carlo simulation, in the dijet invariant mass region of [45,160]~GeV/$c^2$ we predict 716 signal ($S$) events and 29,093 background ($B$) events. The estimated signal fraction ($S/(S+B)$) is small (0.024). The statistical significance ($S/\sqrt{(S+B)}$) is equal to 4.1. In order to increase the sensitivity to $WW$ and $WZ$ in the lepton plus jets final state, more sophisticated techniques beyond event counting are required. Correlations between kinematic quantities are exploited using a feed-forward artificial neural network (ANN) \cite{NN}. 

A feed-forward ANN can be thought of as a single-valued function of input vectors. The function has many parameters, the values of which determine the output for a given input vector. Usually the output is a continuous distribution in the range 0 to 1. The training of the network is equivalent to a minimization procedure. The aim is to reduce the error function, which is the sum of the squared deviations of the neural network output from the desired output for signal (usually 1) and background (usually 0). When the trained network with its optimised parameters is used with real events, the network output for each event is used to define if the event is selected or not \cite{cdfStat}.

The ANN we developed for this analysis is trained using six input variables that can discriminate the signal from the background. The output of the ANN is a variable where the signal and background are well separated. We perform the ANN training using angles and event shape quantities and ensure that cutting on the output of this ANN does not introduce significant bias on the signal and background dijet invariant mass shapes.
 
\begin{figure}[h]
  \begin{center}
    \includegraphics[trim = 10mm 20mm 13mm 20mm, clip, width=0.48\textwidth]{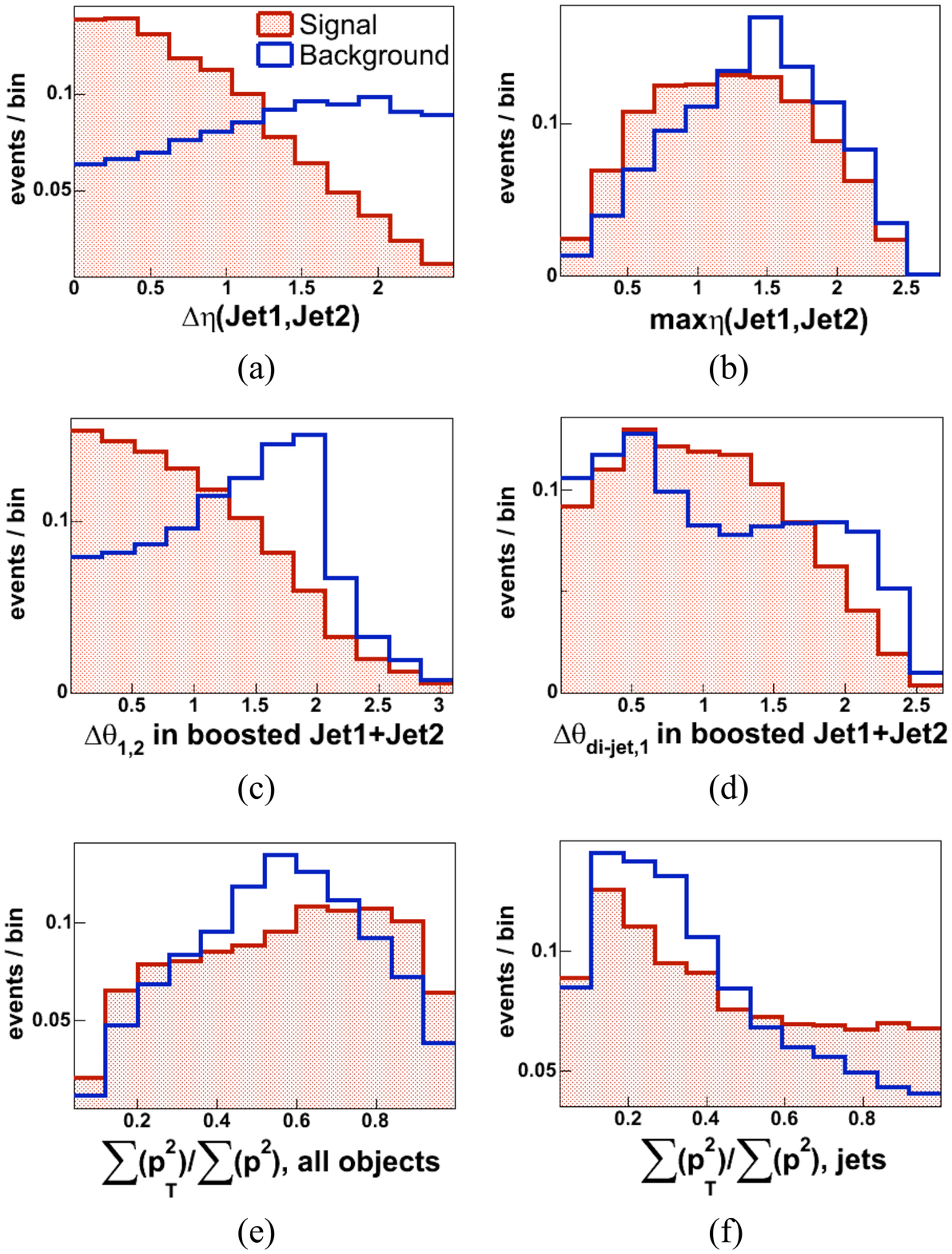} 
    \caption{Neural Network input variables. The ANN is trained with events in the signal region only and for electrons and muons simultaneously. Both signal and background descriptions are given by Monte Carlo simulated events.}
    \label{FIG:nn_in}
  \end{center}
\end{figure}

The quantities used in the ANN training are shown in Fig. \ref{FIG:nn_in}. We used the difference between the pseudorapidity $\rm{\eta}$ of the leading jets ($\Delta\rm{\eta}(Jet1,Jet2)$); the maximum value of the pseudorapidity $\rm{\eta}$ of the two jets ($max~\rm{\eta}(Jet1,Jet2)$); the fraction $\sum p_{T}^{2}/\sum p^{2}$, where the sum is over all objects, leptons and jets (for the neutrino the $\Met$ is used in the denominator); the fraction $\sum p_{T}^{2}/\sum p^{2}$, where the sum is over the two leading jets; and finally the quantities $\Delta\theta_{1,2} = \theta_{Jet1} - \theta_{Jet2}$ and $\Delta\theta_{dijet,1}$, both calculated in the rest frame of the dijet system. The quantity $\Delta\theta_{dijet,1}$ is given by the expression $\Delta\theta_{dijet,1}=|\theta_{Jet1}-\theta_{dijet}|$, if $\theta_{Jet1}\cdot\theta_{dijet}>0$; otherwise, it is given by $\Delta\theta_{dijet,1}=|\pi-\theta_{Jet1}-\theta_{dijet}|$, where $\theta_{Jet1}$ and $\theta_{dijet}$ are calculated in the rest frame of the dijet system.

The training is performed using the variables in the signal region only. Both signal and background descriptions are given by Monte Carlo simulations. The ANN has been trained for the electron and the muon channels combined. 
      
\begin{figure}[h]
  \begin{center}
    \includegraphics[trim = 18mm 20mm 20mm 30mm, clip, width=0.48\textwidth]{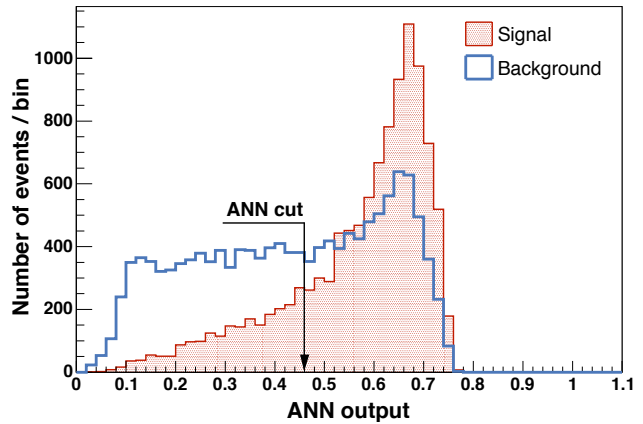} 
    \caption{ANN Output, interpreted as a function associating the output, from 0.0 to 1.0, to each event. The distributions for signal and background samples are shown. A cut is applied at the value 0.46. This is the value where the statistical significance is maximized, in the context of this specific ANN output.}
%        \begin{picture}(0,20)
%             \Text(0.0,60.0)[c]{\footnotesize{\textbf{ANN Output}}}  
%        \end{picture}
    \label{FIG:nn_out}
  \end{center}
\end{figure}

The ANN output is shown in Fig. \ref{FIG:nn_out}. A cut is applied at the point where the statistical significance is maximized. After applying this cut in Monte Carlo simulated events, we estimate within the dijet invariant mass region of [45,160]~GeV/$c^2$ an expected number of 554 signal events and 14,481 background events. The signal fraction is 0.037, improved by 53\% with respect to the value before the ANN cut was applied. The statistical significance is 4.5, an improvement of about 10\%.

\begin{figure}[h]
 
    \begin{center}
      \includegraphics[trim=15mm 16mm 20mm 28mm, clip,width=0.48\textwidth]{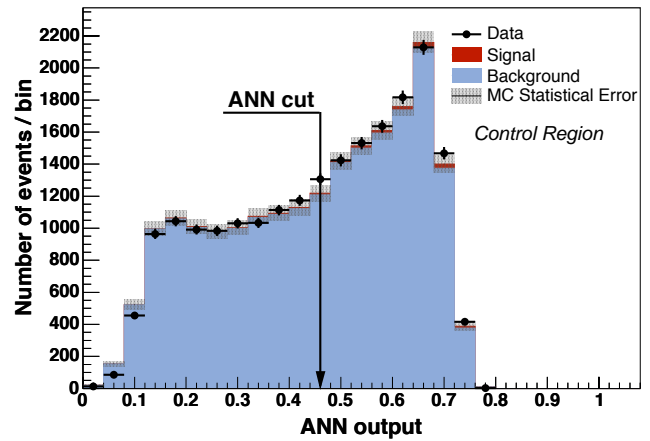} 
      \caption{Comparison of the experimental data with Monte Carlo simulation in the dijet invariant mass sidebands for the ANN output. The $\chi^2$ probability the agreement between the experimental data and the Monte Carlo simulation is $\approx30\%$.}
      \label{FIG:NN_out}
    \end{center}
\end{figure}

A comparison between the data and the Monte Carlo simulated events of the ANN output shape in the sideband regions is shown in Fig. \ref{FIG:NN_out}. For the ANN output in the sidebands, the data are well described by the Monte Carlo simulation.

We measure the signal fraction in the data by performing a likelihood fit on the dijet invariant mass distribution. The shape of the dijet invariant mass is parameterized for the events that pass the cut in the ANN output. To obtain the signal parameterization, we use {\sc pythia} Monte Carlo simulated events. The background model is derived from the Monte Carlo simulation and is fit with the form ${\rm PDF}_{BGR}\propto \exp{(\alpha x + \beta x^2)}$ (PDF $\equiv$ Probability Density Function), with $\alpha$ and $\beta$ as free parameters. The overall parameterization consists of the signal and background descriptions, with the signal fraction $f_S$ being one additional free parameter. A likelihood function $\cal{L}$ is constructed using this parameterization with a total of three free parameters $\alpha$, $\beta$, and $f_S$; a fit is performed to the data. The fit returns the parameters $\alpha$ and $\beta$ as well as the signal fraction, which is then converted into the number of events.

The signal fraction measured from the data over a dijet invariant mass region of [45,160]~GeV/$c^2$, and for the events that pass the ANN cut, is $f_S=0.027\pm0.014$.
Given a total of 15,016 events, this signal fraction corresponds to $410\pm213$ signal events. The uncertainty is statistical, obtained from the fit, and accounts for the poisson fluctuations of the total number of events measured on the data.  

\begin{figure}[h]
  \centering 
  \includegraphics[trim=15mm 10mm 30mm 10mm, width=.48\textwidth]{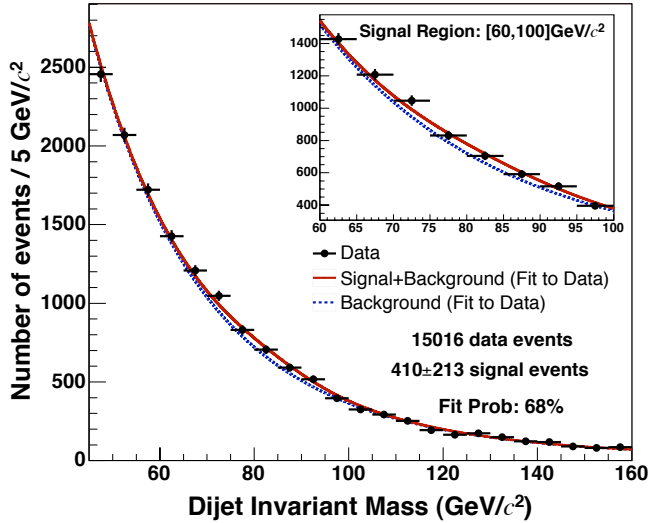} 
  \caption{Likehood fit on data (solid line). The dashed line shows the background estimation, as given by the data. The inset provides a close-up in the signal region.}
  \label{FIG:Fit1} 
\end{figure}

\begin{figure}[h]
  \centering 
  \includegraphics[trim=20mm 10mm 30mm 25mm, width=.48\textwidth]{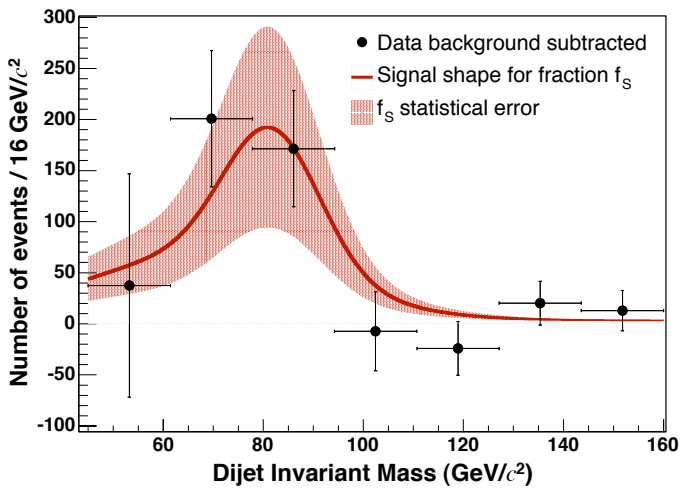} 
  \caption{The signal shape measured on data. The plot is made by subtracting the measured background shape from the data. The errors are statistical only. The solid line corresponds to the Monte Carlo simulated events shape for the measured signal fraction. There is a good agreement between the Monte Carlo events shape and the shape seen in the data.}
  \label{FIG:Fit3} 
\end{figure}

The likelihood fit from the data is shown in Fig. \ref{FIG:Fit1}. The overall fit result (signal plus background) and the measured background shape are displayed. Figure \ref{FIG:Fit3} shows the signal shape that is measured on the data, obtained by subtracting the background from the data. The signal shape is compared to the expected signal shape, normalized according to the measured signal fraction.

The significance of the result is evaluated using the likelihood ratio, ${\cal Q} = {\cal L}_{S+B}(\alpha,\beta,f_S)/{\cal L}_{B}(\alpha,\beta)$, as a test statistic.  We test the SM signal plus background hypothesis and the background-only hypothesis by analyzing a set of simulated experiments, as done in data. We perform a fit for the three free parameters, and each time calculate the likelihood ratio. Using Monte Carlo simulated experiments, we estimate a $\approx$ 2.5$\sigma$ statistical significance for the expected signal given its SM cross section. From the data, we measure a 1.9$\sigma$ statistical significance. The data are thus compatible with SM expectations, and we estimate an upper limit of the $WW$ and $WZ$ cross section.

%%\section{Cross section}

\vspace{0.5cm}

The cross section times branching ratio that corresponds to the measured number of signal events is estimated using the formula $\sigma \times BR = N_{signal} / \alpha \cdot \epsilon \cdot {\mathscr L} $, where $N_{signal}$ is the measured number of signal events; $\alpha$ is the signal acceptance, derived from the Monte Carlo simulated events; $\epsilon$ is the global efficiency factor that includes vertex, tracking and trigger efficiencies; and ${\mathscr L}$ is the total integrated luminosity of the data we used. The product of the acceptance times efficiency is estimated separately for electrons and muons.

%%\section{Systematics}

To assess the effects of systematic uncertainties on the measurement, we address separately two kinds of systematic uncertainties: those that affect the signal fraction, and those that affect the acceptance. The signal fraction uncertainties define the uncertainty in the significance of the measurement. 

The dominant systematic uncertainty in the signal fraction come from the background shape parameterization. The background shape is fit to the form ${\rm PDF}_{BGR}\propto
\exp{(\alpha x + \beta x^2)}$, as already described, which has two parameters $\alpha$ and $\beta$. This form gives an adequate fit to both the Monte Carlo simulation and the data. In order to quantify the size of the systematic uncertainty associated with the background shape, fits with additional parameters in the exponent (from three to six) were carried out. The variations obtained were used to assign the systematic uncertainty.

Other systematic uncertainties in the signal fraction include those originating from the energy scale of the jets (JES). The effect in the signal fraction is quantified by varying the parameters of the signal shape to account for the $\pm 1 \sigma$ variations of the JES. We generate simulated experiments with the new parameterizations and fit them with the standard signal parameterization. The difference in the signal fraction from the different fits determines the systematic uncertainty. 

The result also depends on the dijet invariant mass resolution which in turn depends on the jet energy resolution. To estimate the systematic uncertainty due to this effect, we introduce an additional Gaussian smearing relative to the jet energy. Other systematic uncertainties that affect the signal shape but have smaller effects on the result are the initial and final state radiation effects.

A summary of all systematic uncertainties on the signal fraction and their effect on the measurement is given in Table \ref{table_sys}. The total systematic uncertainty on the signal fraction is estimated to be 25\%. 

The systematic uncertainties affecting the acceptance are evaluated by counting the number of events that pass the selection cuts, after varying the various uncertainty sources. The sources that have been taken into account, as well as the actual effect on the acceptance are listed in Table \ref{table_sys}.

The overall uncertainty on the cross section is given by taking into account the uncertainties in the signal fraction, the acceptance and the luminosity. The total effect is estimated to be 26\%. The total uncertainty in the measurement is given by the statistical and systematic uncertainties added in quadrature. Taking into account the systematic uncertainties, the significance of the measurement is 1.7$\sigma$.

\begin{table}[h]
  \begin{center}
    \begin{tabular}{c c}
      \hline

Source & Signal Fraction \\ \hline \hline
Jet Energy Scale & $10\%$ \\ \hline
Jet Resolution     &  $10\%$ \\ \hline
Background Shape &  $20\%$ \\ \hline
Initial State Radiation &  $5\%$ \\ \hline
Final State Radiation & $<1\%$ \\ \hline \hline
Source & Acceptance \\ \hline \hline
Jet Energy Scale & $3\%$ \\ \hline
Jet Resolution     &  $<1\%$ \\ \hline
Initial State Radiation &  $2\%$ \\ \hline
Final State Radiation & $3\%$ \\ \hline
Efficiency Factor & $3\%$ \\ \hline \hline
Source & Cross Section \\ \hline \hline
Total Signal Fraction & $25\%$ \\ \hline
Total Acceptance & $5\%$ \\ \hline
Luminosity & $6\%$ \\ \hline \hline
Total Effect in Cross Section & $26\%$ \\ \hline
      \hline
    \end{tabular}
  \end{center}
  \caption{The systematic uncertainties and their effect in the signal fraction, the acceptance and finally the cross section. All uncertainties are added in quadrature.} 
  \label{table_sys}

\end{table}

\vspace{0.5cm}
Taking into account both the statistical and systematic error in the number of signal events, we measure
%\begin{equation}
%\begin{split}
$\sigma_{WW}\times BR(W\rightarrow \ell\nu_{\ell},W\rightarrow \rm{jets})+\sigma_{WZ}\times BR(W\rightarrow \ell\nu_{\ell},Z\rightarrow \rm{jets} ) =
1.47 \pm0.77(stat)\pm0.38(sys)~\rm{pb}$, which is consistent with the SM theoretical prediction for the cross section, $2.09\pm0.12~\rm{pb}$.
We set a 95$\%$ confidence level (C.L.) upper limit for the measured cross section. Given that the uncertainties follow a Gaussian distribution, the 95$\%$ C.L. limit can be set by the estimated value plus 1.65 standard deviations \cite{stat_data_analysis}. The 95\% C.L. upper limit set for the cross section is $\sigma\times BR<2.88~\rm{pb}$.

\vspace{0.1cm}

In summary, we have used 1.2~fb$^{-1}$ of CDF II data to search for $WW$ and $WZ$ production in the lepton plus jets final state. We use an ANN to discriminate the signal from the background. This technique improves the expected statistical significance by $\approx10\%$ and the expected signal fraction by 50\%, as compared with an event 
selection based on conventional cuts. We find no evidence for anomalous $WW$ and $WZ$ production in the lepton plus jets final state and we set a 95\% C.L. upper limit for the cross section at $\sigma\times BR<2.88~\rm{pb}$.

\vspace{0.1cm}

We thank the Fermilab staff and the technical staffs of the participating institutions for their vital contributions. This work was supported by the U.S. Department of Energy and National Science Foundation; the Italian Istituto Nazionale di Fisica Nucleare; the Ministry of Education, Culture, Sports, Science and Technology of Japan; the Natural Sciences and Engineering Research Council of Canada; the National Science Council of the Republic of China; the Swiss National Science Foundation; the A.P. Sloan Foundation; the Bundesministerium f\"ur Bildung und Forschung, Germany; the Korean Science and Engineering Foundation and the Korean Research Foundation; the Science and Technology Facilities Council and the Royal Society, UK; the Institut National de Physique Nucleaire et Physique des Particules/CNRS; the Russian Foundation for Basic Research; the Ministerio de Ciencia e Innovaci\'{o}n, and Programa Consolider-Ingenio 2010, Spain; the Slovak R\&D Agency; and the Academy of Finland. 

% ======================================================================

\end{document}